\documentclass[10pt,journal,compsoc]{IEEEtran}
\usepackage{xcolor,soul,framed} 
\PassOptionsToPackage{hyphens}{url}\usepackage{hyperref}
\colorlet{shadecolor}{yellow}
\usepackage[pdftex]{graphicx}
\graphicspath{{../pdf/}{../jpeg/}}
\DeclareGraphicsExtensions{.pdf,.jpeg,.png}

\usepackage{amsmath}
\usepackage{amssymb}
\usepackage{mathabx}
\usepackage{array}
\usepackage{mdwmath}
\usepackage{mdwtab}
\usepackage{eqparbox}
\usepackage{url}

\usepackage{algorithm}
\usepackage{algorithmic}

\usepackage{multirow}
\usepackage{subfig}
\usepackage{tablefootnote}

\ifCLASSOPTIONcompsoc
  \usepackage[nocompress]{cite}
\else
  \usepackage{cite}
\fi
\ifCLASSINFOpdf
\else
\fi
\begin{document}
%
\title{Cutting-edge 3D Medical Image Segmentation Methods in 2020: Are Happy Families All Alike?}
%
%
%
%

\author{Jun~Ma
\IEEEcompsocitemizethanks{\IEEEcompsocthanksitem Jun Ma was with the Department
of Mathematics, School of Science, Nanjing University of Science and Technology, Nanjing, China,
210094.\protect\\
}
\thanks{Manuscript was finished on December 31, 2020. Comments are welcome.}}

\IEEEtitleabstractindextext{%
\begin{abstract}
Segmentation is one of the most important and popular tasks in medical image analysis, which plays a critical role in disease diagnosis, surgical planning, and prognosis evaluation.
During the past five years, on the one hand, thousands of medical image segmentation methods have been proposed for various organs and lesions in different medical images, which become more and more challenging to fairly compare different methods.
On the other hand, international segmentation challenges can provide a transparent platform to fairly evaluate and compare different methods.
In this paper, we present a comprehensive review of the top methods in ten 3D medical image segmentation challenges during 2020, covering a variety of tasks and datasets.
We also identify the "happy-families" practices in the cutting-edge segmentation methods, which are useful for developing powerful segmentation approaches. Finally, we discuss open research problems that should be addressed in the future.
We also maintain a list of cutting-edge segmentation methods at \url{https://github.com/JunMa11/SOTA-MedSeg}.
\end{abstract}

}

\maketitle

\IEEEdisplaynontitleabstractindextext

%
\IEEEpeerreviewmaketitle

\IEEEraisesectionheading{\section{Introduction}\label{sec:introduction}}

\begin{figure}[htbp]
\centering
\includegraphics[scale=0.3]{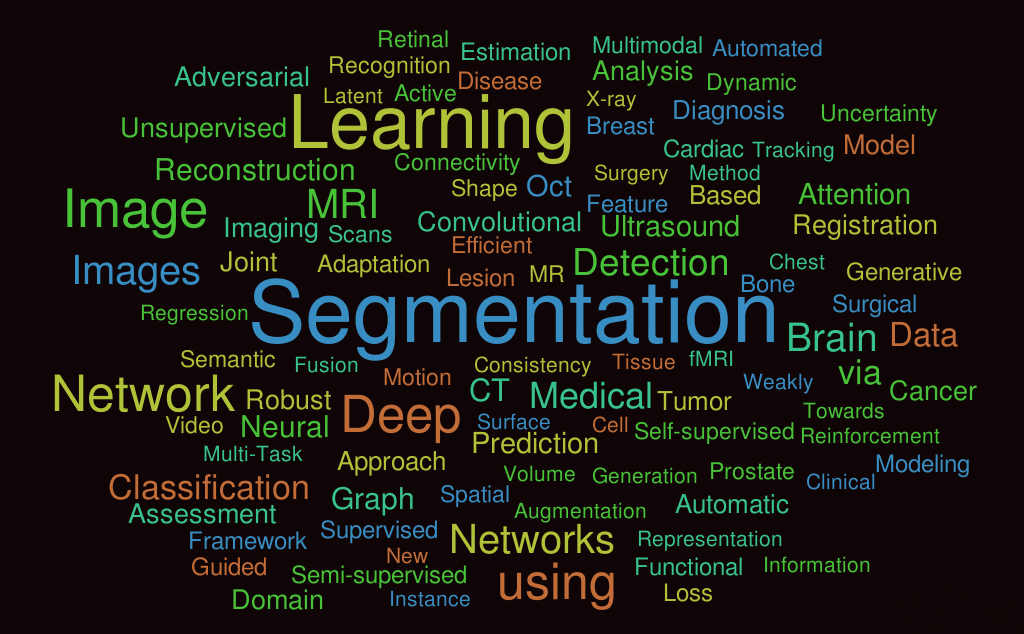}
\caption{Word cloud of the paper titles in MICCAI 2020.}\label{intro-wordcloud}
\end{figure}

%
%
%
%
\IEEEPARstart{M}{edical} image segmentation aims to delineate the interested anatomical structures, such as tumors, organs, and tissues, in a semi-automatic or fully automatic way, which has many applications in clinical practice, such as radiomic analysis~\cite{gillies2016radiomics}, treatment planning~\cite{rietzel2005treatmentplanning}, and survival analysis~\cite{zhang2020COVIDCell}, and so on.
Currently, medical image segmentation is also an active research topic. Figure~\ref{intro-wordcloud} presents the word cloud of the paper titles in the 23rd International Conference and Medical Image Computing \& Computer Assisted Intervention (MICCAI 2020)\footnote{https://miccai2020.org/en/} that is the largest international event in medical image analysis community.
It can found that the term `segmentation' has very high frequency and putting the top high-frequency words together can form a meaningful phase ``image segmentation using deep learning/network(s)''.

Since U-Net~\cite{ronneberger2015UNet2D}, the legend medical image segmentation approach, appeared in 2015, there has been numerous new segmentation methods have been proposed for various segmentation tasks~\cite{litjens2017MIA-DL-survey, Boykov2020DLSeg} in the past five years\footnote{U-Net~\cite{ronneberger2015UNet2D} has more than 20000 citations by December, 2020.}.
With so many segmentation papers on hand, it becomes extremely hard to compare them and identify the methodology progress, because the proposed methods are usually evaluated on different datasets with different dataset splits, metrics, and implementations.

Public segmentation challenges provide a standard platform of getting insights into the current cutting-edge approaches where solutions are evaluated and compared against each other in a transparent and fair way.
In MICCAI 2020, there are totally ten international 3D medical image segmentation challenges\footnote{https://www.miccai2020.org/en/MICCAI-2020-CHALLENGES.html}. All these challenges follow the Biomedical Image Analysis ChallengeS (BIAS) Initiative~\cite{maier2020BIAS-MIA}. Specifically, the challenge designs are transparent and standardized, and the proposals (\url{http://miccai.org/events/challenges/}) also have passed the peer review.

Table~\ref{tab:task-overview} provides an overview of the 10 segmentation challenges, which can be roughly divided into
\begin{itemize}
    \item \textbf{five single-modality} image segmentation tasks, including three CT image segmentation tasks and two MR image segmentation tasks;
    \item \textbf{five multi-modality} image segmentation tasks, including two bi-modality tasks, two triple-modality tasks, and one four-modality task.
\end{itemize}

\begin{table*}[!htbp]
\centering
\caption{Task overview of ten 3D medical image segmentation challenges. `Seg. Targets' denotes segmentation targets in each task; `\# Class and \# Train/Val./Test' denote the number of class and the number of cases in training set, validation set, and testing set, respectively. `-' denotes no validation cases. `+' denotes that additional segmentation metrics (except DSC and HD) are used to evaluate the solutions of challenge participants.}
\label{tab:task-overview}
\resizebox{\textwidth}{!}{
\begin{tabular}{llccccl}
\hline
Name                                                                                                     & Seg. Targets                                                           & \# Class & \# Train/Val./Test          & Modality                         & Multi-Center         & Metrics                     \\
\hline
CADA                                                                                                     & Cerebral Aneurysm                                                      & 1        & 92/-/23                     & CT                               &                      & IoU, HD, +                  \\
ASOCA                                                                                                    & Coronary arteries                                                      & 1        & 40/-/20                     & CT                               &                      & DSC, HD                     \\
VerSeg                                                                                                   & Vertebra                                                               & 28       & 100/-/200                   & CT                               & Y                    & DSC, HD                     \\
MMs                                                                                                      & Heart\tablefootnote{Myocardium, left and right ventricle}              & 3        & 150(+25)/-/200              & MR                               & Y                    & DSC, HD, +                  \\
EMIDEC                                                                                                   & Myocardium, infraction, re-flow                                        & 3        & 100/-/50                    & MR                               &                      & DSC, HD, +                  \\ \hline
ADAM                                                                                                     & Intracranial aneurysm                                                  & 1        & 113/-/140                   & TOF-MRA, structural MR           &                      & DSC, HD, +                  \\
HECKTOR                                                                                                  & Head/neck tumor                                                        & 1        & 203/-/46                    & PET, CT                          & Y                    & DSC                         \\
MyoPS                                                                                                    & Scar, edema                                                            & 2        & 25/-/20                     & LGE, T2, bSSFP                   &                      & DSC                         \\
\multirow{2}{*}{ABCs}                                                                                    & Task 1: 5 brain structures                                             & 5        & \multirow{2}{*}{45/15/15}   & \multirow{2}{*}{CT, T1, T2}      & \multirow{2}{*}{Y}    & \multirow{2}{*}{DSC, SDSC}  \\
                                                                                                         & Task 2: 10 brain structures                                            & 10       &                             &                                  &                      &                             \\
BraTS                                                                                                    & Brain tumor\tablefootnote{Whole tumor, enhancing tumor, and tumor core}  & 3        & 369/125/166                 & Flair, T1, T1ce, T2              & Y                    & DSC, HD                     \\
\hline
\end{tabular}}
\end{table*}

In this paper, we first provide a comprehensive review of the ten 3D medical segmentation challenges and the associated top solutions.
we also identify the "happy-families" elements in the top solutions.
Finally, we highlight some problems and potential future directions for medical image segmentation.

The main contributions of this paper are summarized as follows:
\begin{itemize}
    \item We provide a comprehensive review of ten recent international 3D medical image segmentation challenges, including the task descriptions, the datasets, and more importantly, the top solutions of participant teams, which represent the cutting-edge segmentation methods at present.
    \item We identify the widely used "happy-families" components in the top methods, which are useful for developing powerful segmentation approaches.
    \item We summarize several unsolved problems and potential research directions, which could promote the developments in medical image segmentation field.
\end{itemize}

\section{Preliminaries: Widely Used Methods in Deep Learning-based Medical Image Segmentation}

\subsection{Network Architectures}
nnU-Net~\cite{isensee2020nnunet}, no new net, is a dynamic fully automatic segmentation framework for medical images, which is based on the widely used U-Net architecture~\cite{ronneberger2015UNet2D, cciccek2016UNet3D}.
It can automatically configures the preprocessing, the network architecture, the training, the inference, and the post-processing for any new segmentation task.
Without manual intervention, nnU-Net surpasses most existing approaches, and achieves the state-of-the-art in 33 of 53 segmentation tasks and otherwise shows comparable performances to the top leaderboard entries.
Currently, nnU-Net has been the most popular backbone for 3D medical image segmentation tasks because of its powerful, flexible, out-of-the-box, and open-sourced,

\subsection{Loss Functions}
Loss function is used to guide the network to learn meaningful predictions and dictate how the network is supposed to trade off mistakes. Cross entropy loss and Dice loss~\cite{milletari2016vnet, diceV2} are two most popular loss functions in segmentation tasks. Specifically, cross entropy aims to minimize the dissimilarity between two distributions, which is defined by
\begin{equation}
    L_{CE} = -\frac{1}{N}\sum_{c=1}^{C}\sum_{i=1}^{N} g_{i}^{c} \log s_{i}^{c},
\end{equation}
where $g_i^c$ is the ground truth binary indicator of class label $c$ of voxel $i$, and $s_i^c$ is the corresponding predicted segmentation probability.

Dice loss can directly optimize the Dice Similarity Coefficient (DSC) which is the most commonly used segmentation evaluation metric. In general, there are two variants for Dice loss, one employs squared terms in the denominator \cite{milletari2016vnet}, which is defined by
\begin{equation}\label{eq:DiceV1}
    L_{Dice-square} = 1- \frac{2\sum_{c=1}^{C}\sum_{i=1}^{N}g_{i}^{c}s_{i}^{c}}{\sum_{c=1}^{C}\sum_{i=1}^{N}(g_{i}^{c})^2 + \sum_{c=1}^{C}\sum_{i=1}^{N}(s_i^{c})^2}.
\end{equation}
The other does not use the squared terms in the denominator \cite{diceV2}, which is defined by
\begin{equation}\label{eq:Dice}
    L_{Dice} = 1- \frac{2\sum_{c=1}^{C}\sum_{i=1}^{N}g_{i}^{c}s_{i}^{c}}{\sum_{c=1}^{C}\sum_{i=1}^{N}g_{i}^{c} + \sum_{c=1}^{C}\sum_{i=1}^{N}s_i^{c}}.
\end{equation}
The default loss function in nnU-Net is the unweighted sum $L_{CE} + L_{Dice}$.

\subsection{Evaluation Metrics}
Dice Similarity Coefficient (DSC) and Hausdorff Distance (HD) are two widely used segmentation metrics, which can measure the region overlap ratio and boundary distance, respectively.
Let $G$ and $S$ be the ground truth and the segmentation result, respectively.
DSC is defined by
\begin{equation}
    DSC = \frac{2|G\cap S|}{|G|+ |S|}.
\end{equation}
A similar metric IoU (Jaccard) sometimes is used as an alternative, which is defined by
\begin{equation}
    IoU = \frac{|G\cap S|}{|G\cup S|}.
\end{equation}

Let $\partial G$ and $\partial S$ are the boundary points of the ground truth and the segmentation, respectively. The Hausdorff Distance is defined by
\begin{equation}
    HD(\partial G, \partial S) = \max(hd(\partial G, \partial S), hd(\partial S, \partial G)),
\end{equation}
where
\begin{equation*}
    hd(\partial G, \partial S) = \max\limits_{x\in \partial G} \min\limits_{y\in \partial S} ||x-y||_2,
\end{equation*}
and
\begin{equation*}
    hd(\partial S, \partial G) = \max\limits_{x\in \partial S} \min\limits_{y\in \partial G} ||x-y||_2.
\end{equation*}
To eliminate the impact of the outliers, 95\% HD is also widely used, which is based on the calculation of the 95th percentile of the distances between boundary points in $\partial G$ and $\partial S$.

\section{Single Modality Image Segmentation}

\begin{figure*}[htbp]
\centering
\includegraphics[scale=0.8]{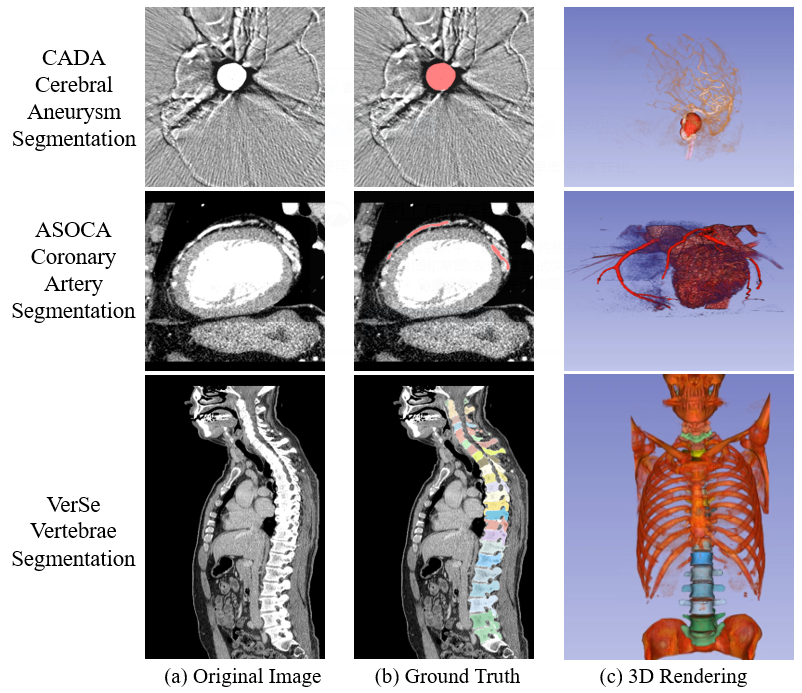}
\caption{Visualized examples in three CT segmentation tasks. The ground truth of the original image (a) in each task is shown in 2D projected onto the raw data (b) and in 3D together with a volume rendering of the raw data (c).
}\label{fig:CT-Seg}
\end{figure*}

\subsection{CADA: Cerebral Aneurysm Segmentation}
The task in CADA challenge (\url{https://cada-as.grand-challenge.org/Overview/}) is to segment the aneurysms from 3D CT images.
The organizers provide 92 cases for training and 23 cases for testing, where the cases are with cerebral aneurysms without vasospasm. The main difficulty in this challenge is the highly imbalanced labels. As shown in Figure~\ref{fig:CT-Seg} (the first row), the aneurysm is very small and most of the voxels are the background in the CT images.

Six metrics are used to quantitatively evaluate the segmentation results, including Jaccard (IoU), Hausdorff distance (HD), mean distance (MD), Pearson correlation coefficient between predicted volume and reference volume of all aneurysms (Volume Pearson R), the mean absolute difference of predicted and reference volume  (Volume Bias), and Standard deviation of the difference between predicted and reference volumes (Volume Std).
For the ranking, a maximum-minimum normalization is performed according to all participants. In this way, each individual metric takes a value between 0 (worst case among all participants) and 1 (perfect fit between the reference and predicted segmentation). The ranking score is calculated as the average of the normalized metrics.

\begin{table}[!htbp]
\caption{Quantitative results of top-2 teams on CADA Challenge Leaderboard. The bold numbers denote the best results.}\label{tab:CADA-Results}
\centering
\begin{tabular}{lcc}
\hline
\multirow{2}{*}{Metrics} & Mediclouds     & junma~\cite{Ma20-CADA-2nd}          \\ \cline{2-3}
                         & Rank 1st       & Rank 2nd            \\ \hline
IoU                      & 0.758          & \textbf{0.759}          \\
HD                       & \textbf{2.866}          & 4.967          \\
MD                       & \textbf{1.618}          & 3.535          \\
Vol. Pearson R           & \textbf{0.998}          & \textbf{0.998}          \\
Vol.Bias                 & \textbf{72.24}          & 75.84          \\
Vol.Std                  & \textbf{106.4}          & 110.5          \\ \hline
\textbf{Final Score}    & \textbf{0.833} & 0.832 \\ \hline
\end{tabular}
\end{table}

Table~\ref{tab:CADA-Results} shows the quantitative segmentation results of the top-2 teams on the challenge leaderboard\footnote{\url{https://cada-as.grand-challenge.org/FinalRanking/}}.
The team `junma' achieved the best IoU while the team `Mediclouds' achieved better performance in the remaining five metrics. However, the final score difference is marginal.
The method of the team `Mediclouds', unfortunately, is not available. Thus, we only present the solution of the team `junma'.
Specifically, Ma and Nie~\cite{Ma20-CADA-2nd} developed their methods based on nnU-Net~\cite{isensee2020nnunet} where the main modification to use a large patch size ($192\times224\times192$) during training and inference. Five models were trained in five-fold cross-validation and each model was trained on a TITAN V100 32G GPU.
Each testing case is predicted by the ensemble of the trained five models.

\subsection{ASOCA: Automated Segmentation Of Coronary Arteries}
The task in ASOCA challenge (\url{https://asoca.grand-challenge.org/Home/}) is to segment the coronary arteries from Cardiac Computed Tomography Angiography (CCTA) images. The organizers provide 40 cases for training and 20 cases for testing. The main difficulties in this challenge are the imbalanced problem and appearance variations. On the one hand, the coronary arteries only occupy a small proportion in the whole CT image. On the other hand, the arteries from healthy and unhealthy cases share different appearances.  Figure~\ref{fig:CT-Seg} (the second row) presents a visualized example.
DSC and HD95 are used to evaluate and rank the segmentation results.

\begin{table}[!htbp]
\caption{Quantitative results of top-2 teams on ASOCA Challenge Leaderboard. The bold numbers denote the best results.}\label{tab:ASOCA-Results}
\centering
\begin{tabular}{lccc}
\hline
Team Name      & DSC   & HD95  & Final Rank \\ \hline
RuochenGao & \textbf{0.867} & 4.165 & 1          \\
SenYang    & 0.838 & \textbf{2.336} & 2          \\ \hline
\end{tabular}
\end{table}

Table~\ref{tab:ASOCA-Results} shows the quantitative segmentation results of the top-2  teams on the challenge leaderboard\footnote{\url{https://asoca.grand-challenge.org/MICCAI_Ranking/}} during MICCAI 2020.
The 1st-place team had better DSC while the 2nd-place team obtained better HD95, indicating that the top-2 teams achieved better region overlap and boundary distance, respectively.

The team `RuochenGao' used nnU-Net~\cite{isensee2020nnunet} as the backbone. The whole pipeline include three independent networks for three tasks: epicardium segmentation, artery segmentation, and scale map regression~\cite{wang2020DTM-CVPR}. The final segmentation results were generated by the ensemble of artery segmentation results and scale map regression results followed by removing the vessels outside the epicardium.
The team `SenYang' proposed an improved 2D U-Net with selective kernel (SK-UNet) where the regular convolution blocks were replaced by SE-Res modules in the encoder. Moreover, the SK-modules~\cite{li2019selectivekernel}, including different convolution filters and kernel sizes, were used in the decoder to leverage multi-scale information.

\subsection{VerSe: Large Scale Vertebrae Segmentation Challenge}
The segmentation task in VerSe challenge (https://verse2020.grand-challenge.org/) is to segment the vertebrae from CT images.
The organizers provide 100 cases for training, 100 cases for public testing (the participants can access the testing cases) and 100 cases for hidden testing (this testing set is not publicly available and participants are required to submitted their solutions with Docker containers)~\cite{verseg2, verseg3}. The annotations consist of 28 different vertebrae but each case may only contain part of the vertebrae.
There are several difficulties in this challenge: highly varying fields-of-view (FoV) across cases, large scan sizes, highly correlating shapes of adjacent vertebrae, scan noise, the presence of vertebral fractures, metal implants, and so on~\cite{verseg-benchmark}. Figure~\ref{fig:CT-Seg} (the third row) presents a visualized example.
DSC and HD are used to evaluate and rank the segmentation results.

Payer et al., the defending champion in VerSe 2019~\cite{verseg-benchmark}, succeeded in winning this year's challenge again by the SpatialConfiguration-Net~\cite{payer2019MIA} and U-Net~\cite{ronneberger2015UNet2D,cciccek2016UNet3D}.
Specifically, they proposed a coarse-to-fine approach, including three stages:
\begin{itemize}
    \item stage 1: localizing the whole spine by a 3D U-Net-based heatmap regression network, which can remove background; The network input size ranged from $32\times32\times32$ to $128\times128\times128$.
    \item stage 2: localizing and identifying all vertebrae landmarks simultaneously via a 3D SpatialConfiguration-Net, which combines local appearance with spatial configuration of landmarks; The network input size ranged from $64\times64\times64$ to $96\times96\times256$ during training and was up to $128\times128\times448$ during inference. To address the missed vertebrae, a MRF-based graphical model was employed to refine the localization results.
    \item stage 3: segmenting each vertebra individually by a 3D U-Net. The input size was $128\times128\times96$.
\end{itemize}
Table~\ref{tab:verse} presents the quantitative segmentation results on the public testing set.
The absent results will be added when the challenge summarize paper is released.

\begin{table}[!htbp]
\caption{Quantitative vertebrae segmentation results of the winner solution in VerSe 2020. `-' denotes not available currently.}\label{tab:verse}
\centering
\begin{tabular}{lcc}
\hline
Testing set & DSC    & HD95 \\ \hline
Public      & 0.9354 & -    \\
Hidden      & -      & -    \\ \hline
\end{tabular}
\end{table}

\begin{figure*}[!htbp]
\centering
\includegraphics[scale=0.8]{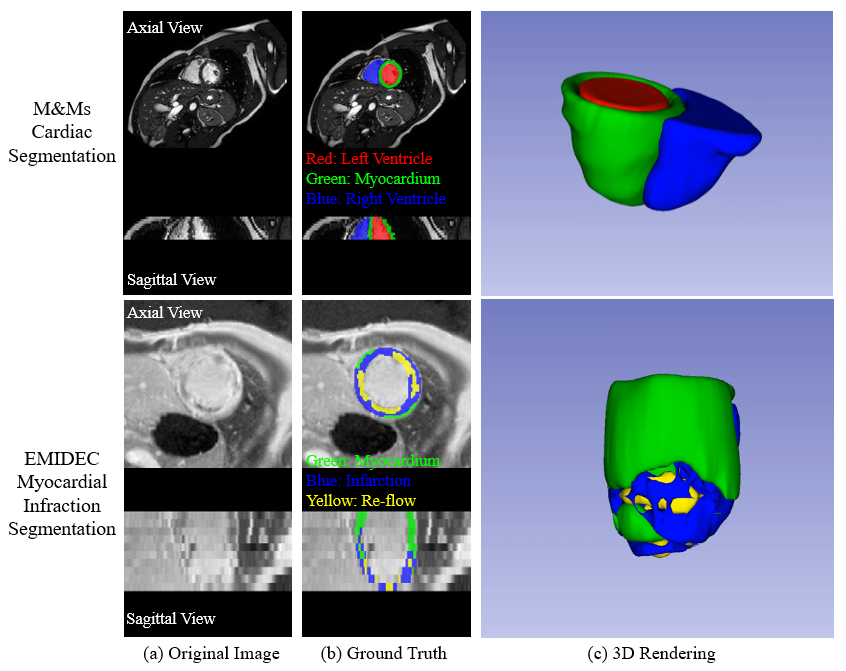}
\caption{Visualized examples in two MR segmentation tasks. The ground truth of the original image (a) in each task is shown in 2D projected onto the raw data (b) and in 3D rendering (c).
}\label{fig:MR-Seg}
\end{figure*}

\subsection{M\&Ms: Multi-Centre, Multi-Vendor \& Multi-Disease Cardiac Image Segmentation Challenge}
The task in M\&Ms challenge \url{https://www.ub.edu/mnms/} is to segment the left and right ventricle (LV and RV, respectively) cavities and the left ventricle myocardium (MYO) from multi-center, multi-vendor, and multi-disease cardiac MR images.
The organizers provide 175 cases for training, 40 cases for validation, and 160 cases for testing, which are from four scanner vendors. Specifically, the 175 training cases consist of 75 labelled cases from the vendor A, 75 labelled cases from the vendor B, and 25 unlabelled cases from the vendor C. The 40 validation cases consist of 10 cases from each of the four vendors. The 160 testing cases consist of 40 cases from each of the four vendors. The main difficulty in this challenge is the domain shift in testing set, which requires the solutions should be generalizable across different clinical centers, scanner vendors, and patient conditions.
It should be noted that both validation cases and testing cases are not publicly available to participants during the challenge. Participants are required to build a Singularity image and shared it with the organizers.
Figure~\ref{fig:MR-Seg} (the first row) presents a visualized example.
Four metrics are used to evaluate and rank the segmentation results, including DSC, IoU, Average symmetric surface distance (ASSD), and HD.

\begin{table}[!htbp]
\caption{Quantitative segmentation results (in terms of DSC and HD) of top-3 teams on M\&Ms Challenge Leaderboard. `ED' and `ES` denote the end-diastolic and end-systolic phases cardiac MR images. The bold numbers are the best results and the italics numbers are not-significant when compared with the best results ($p>0.01$ in T-test).}\label{tab:MMs}
\setlength\tabcolsep{3pt}
\centering
\begin{tabular}{lllccc}
\hline
\multicolumn{3}{l}{\multirow{2}{*}{Metrics}}     & Peter M. Full~\cite{MMs-2020-1st}  & Yao Zhang~\cite{MMs-2020-2nd}      & Jun Ma ~\cite{MMs-2020-3rd}         \\ \cline{4-6}
\multicolumn{3}{l}{}                             & Rank 1st       & Rank 2nd       & Rank 3rd       \\ \hline
\multirow{6}{*}{ED} & \multirow{2}{*}{LV}  & DSC & \textbf{0.939}                                     & \textit{0.938}                                 & \textit{0.935}            \\
                    &                      & HD  & \textbf{9.10}                                       & \textit{9.30}                                   & \textit{9.50}              \\
\cline{2-6}
                    & \multirow{2}{*}{MYO} & DSC & \textbf{0.839}                                     & \textit{0.830}                                 & \textit{0.825}            \\
                    &                      & HD  & \textbf{12.8}                                      & \textit{12.9}                                  & \textit{13.3}             \\
\cline{2-6}
                    & \multirow{2}{*}{RV}  & DSC & \textbf{0.910}                                     & \textit{0.909}                                 & \textit{0.906}            \\
                    &                      & HD  & \textbf{11.8}                                      & \textit{12.3}                                  & \textit{12.3}             \\
\hline
\multirow{6}{*}{ES} & \multirow{2}{*}{LV}  & DSC & \textbf{0.886}                                     & \textit{0.880}                                 & \textit{0.875}            \\
                    &                      & HD  & \textbf{9.10}                                       & \textit{9.50}                                   & \textit{10.5}             \\
\cline{2-6}
                    & \multirow{2}{*}{MYO} & DSC & \textbf{0.867}                                     & \textit{0.861}                                 & \textit{0.856}            \\
                    &                      & HD  & \textbf{10.6}                                      & \textit{10.8}                                  & \textit{11.6}             \\
\cline{2-6}
                    & \multirow{2}{*}{RV}  & DSC & \textbf{0.860}                                     & \textit{0.850}                                 & \textit{0.844}            \\
                    &                      & HD  & \textbf{12.7}                                      & \textit{13.0}                                  & \textit{13.0}             \\
\hline
\end{tabular}
\end{table}

The top-3 teams developed their methods based on nnU-Net~\cite{isensee2020nnunet}. Specifically, Full et al.~\cite{MMs-2020-1st}, the 1st-place team, handled the domain shift problem by an ensemble of five 2D and five 3D nnU-Net models that were trained with the batch normalization and extensive data augmentation, such as random rotation, flipping, gamma correction, multiplicative/additive brightness, and so on.
Zhang et al.~\cite{MMs-2020-2nd}, the 2nd-place team, used label propagation to leverage unlabelled cases and exploited the style transfer to reduce the variance among different centers and vendors. The final solution was one single model without using postprocessing and ensemble.
Ma~\cite{MMs-2020-3rd}, the 3rd-place team, addressed the domain shift problem by enlarging the training set with histogram matching, where new training cases were generated by using histogram matching to transfer the intensity distribution of 25 unlabelled cases to existing labelled cases. The final solution was an ensemble of five 3D nnU-Net models.
Table~\ref{tab:MMs} presents the quantitative segmentation results of the top-3 teams. It can be found that the differences among them were marginal and not statistically significant, indicating that \textit{all (three) roads lead to Rome}.

\subsection{EMIDEC: Automatic Evaluation of Myocardial Infarction from Delayed-Enhancement Cardiac MRI}
The task in EMIDEC challenge (\url{http://emidec.com/}) is to segment the myocardium, the infarction, and the no-reflow areas from delayed-enhancement cardiac MR images. The organizers provide 100 cases for training and 50 cases for testing~\cite{lalande2020EMIDEC-Data}. The main difficulties in this challenge are the low contrast, varied short-axis orientations, heterogeneous appearances of myocardium pathology areas, and unbalanced distribution between normal and pathological cases.
Figure~\ref{fig:MR-Seg} (the second row) presents a visualized example.
The evaluation and ranking metrics include
\begin{itemize}
    \item clinical metrics: the average errors for the volume of the myocardium (in mm3), the volume (in mm3) and the percentage of infarction and no-reflow area;
    \item geometrical metrics: the average DSC for the different areas and Hausdorff distance (in 3D) for the myocardium.
\end{itemize}

Table~\ref{tab:EMIDEC} presents the quantitative segmentation results of the top-3 teams on the final leaderboard\footnote{\url{http://emidec.com/leaderboard}}.
Both Zhang and Ma, the top-2 teams, used a two-stage cascaded framework and developed their methods based on nnU-Net~\cite{isensee2020nnunet}. Specifically, Zhang~\cite{zhang20-EMIDEC-1st} first used a 2D nnU-Net, focusing on the intra-slice information, to obtain a preliminary segmentation, and then a 3D nnU-Net, focusing on the volumetric spatial information, was employed to refine the segmentation results. The 3D nnU-Net took the combination of the preliminary segmentation and original image as the input.
Finally, the scattered voxels in segmentation results were removed in postprocessing step.
Ma~\cite{Ma20-EMIDEC-2nd} used the 2D nnU-Net in the two stages. Firstly a 2D U-Net was used to segment the whole heart, including the left ventricle and the myocardium. Then, the whole heart was cropped as a region of interest (ROI). Finally, a new 2D U-Net was trained to segment the infraction and no-reflow areas in the ROI. The final model was an ensemble of five 2D nnU-Net models in each stage.
Feng et al.~\footnote{\url{http://emidec.com/downloads/papers/paper-24.pdf}} used dilated 2D UNet~\cite{zhou2020ACNN} with rotation-based augmentation, which aim to overcoming the varied short-axis orientations.

\begin{table}[!htbp]
\caption{Quantitative results of top-3 teams on EMIDEC Challenge Leaderboard.}\label{tab:EMIDEC}
\setlength\tabcolsep{3pt}
\centering
\begin{tabular}{llccc}
\hline
\multirow{2}{*}{Targets}    & \multirow{2}{*}{Metrics} & Zhang \cite{zhang20-EMIDEC-1st}    & Ma~\cite{Ma20-EMIDEC-2nd}   & Feng et al. \\ \cline{3-5}
                            &                          & Rank 1st & Rank 2nd & Rank 3rd    \\ \hline
\multirow{3}{*}{Myocardium} & DSC                      & 0.8786   & 0.8628   & 0.8356      \\
                            & Vol. Diff.               & 9258     & 10153    & 15187       \\
                            & HD                       & 13.01    & 14.31    & 33.77       \\ \hline
\multirow{3}{*}{Infarction} & DSC                      & 0.7124   & 0.6224   & 0.5468      \\
                            & Vol. Diff.               & 3118     & 4874     & 3971        \\
                            & Vol. Diff. Ratio         & 2.38\%   & 3.50\%   & 2.89\%      \\ \hline
\multirow{3}{*}{Re-flow}    & DSC                      & 0.7851   & 0.7776   & 0.7222      \\
                            & Vol. Diff.               & 634.7    & 829.7    & 883.4       \\
                            & Vol. Diff. Ratio         & 0.38\%   & 0.49\%   & 0.53\%      \\ \hline
\end{tabular}
\end{table}

The methods of Zhang~\cite{zhang20-EMIDEC-1st} and Ma~\cite{Ma20-EMIDEC-2nd} obtained comparable results for myocardium and re-flow areas, but Zhang achieved significantly better results for infraction, which were 9\% and 17\% higher than the methods of Ma~\cite{Ma20-EMIDEC-2nd} and Feng et al. in terms of DSC.
The major methodology difference is that Zhang used the 3D network in the second stage while Ma and Feng et al. used the 2D network. Thus, one of the possible reasons might be that 3D network can use more image contextual information than 2D network, and also lead to better performance.

\section{Multi-modality 3D Image Segmentation}

\subsection{ADAM: Intracranial Aneurysm Detection and Segmentation Challenge}
\begin{figure}[htbp]
\centering
\includegraphics[scale=0.45]{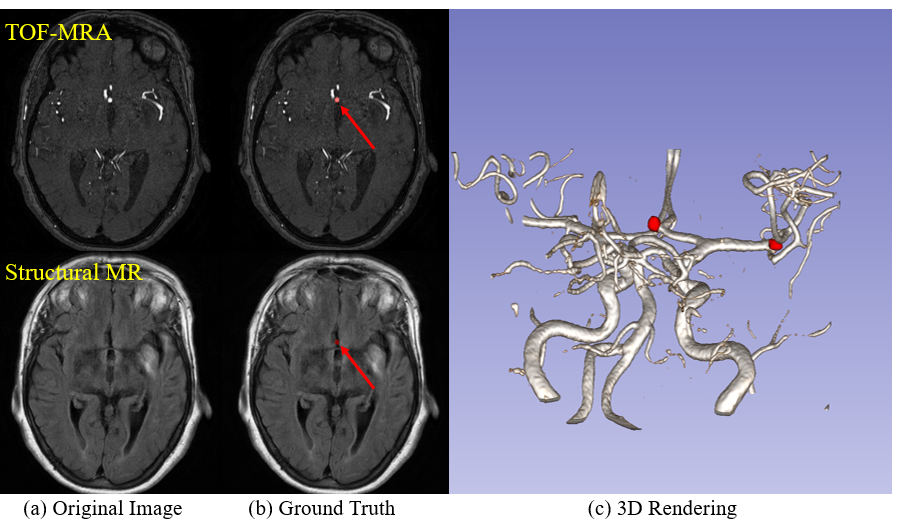}
\caption{Visualized examples in ADAM Challenge. Ground truth (b) of the intracranial aneurysm is shown in 2D projected onto the TOF-MRA and the structural MR image (a) and in 3D together with a volume rendering of the raw data (c). The red arrows point to the intracranial aneurysm.
}\label{fig:ADAM}
\end{figure}

The task in ADAM challenge (\url{http://adam.isi.uu.nl/}) is to segment the aneurysms from TOF-MRA and structural MR images.
The organizers provide 113 cases for training and 141 cases for testing. In the 113 training cases, 93 cases contain at least one untreated, unruptured intracranial aneurysm and 20 cases do not have intracranial aneurysms. In the 141 testing cases, 117 cases containing at least one untreated, unruptured intracranial aneurysm, and 26 cases do not have intracranial aneurysms. Each case has two folders:
\begin{itemize}
    \item `orig' folder: containing all of the original TOF-MRA images and structural images (T1, T2, or FLAIR). The structural image was aligned to the TOF image by elastix\footnote{\url{https://elastix.lumc.nl/}}.
    \item `pre' folder: All images were preprocess by `n4biasfieldcorrection'\footnote{\url{http://stnava.github.io/ANTs/}} to correct bias field inhomogeneities.
\end{itemize}
The main difficulty in this challenge is the extremely imbalanced problem. Specifically, the median image size is $512\times512\times140$, while the median aneurysm voxel size is 238, leading to a extremely imbalanced foreground-background ratio of $6.5\times10^{-6}$.
Figure~\ref{fig:ADAM} presents the visualized examples.
Participants are allowed to use any of the provided images to develop their methods. The testing set is hidden by the organizers and participants should submit their methods with Docker containers.

Table~\ref{tab:ADAM} presents the quantitative results of top-2 teams on ADAM Challenge Leaderboard\footnote{\url{http://adam.isi.uu.nl/results/results-miccai-2020/}} during MICCAI 2020.
Both teams developed their methods based on nnU-Net~\cite{isensee2020nnunet}.
Specifically, to alleviate the imbalanced problem, the team `junma` trained two group five-fold nnU-Net models with Dice + Cross entropy loss and Dice + TopK loss, respectively~\cite{SegLossOdyssey}. Only preprocessed TOF-MRA images were used during training.
The final model was the ensemble of five best models during cross-validation.
To speed up the inference, the default testing time augmentation in nnU-Net (TTA) was disabled during testing.
The team `jocker' modified the default nnU-Net by introducing residual blocks in the encoder and replacing the instance normalization with group normalization. The loss function was Dice + TopK loss. The final model was the ensemble of four models with different modalities and output classes.

\begin{table}[!htbp]
\caption{Quantitative results of top-2 teams on ADAM Challenge Leaderboard. The bold numbers are the best results.}\label{tab:ADAM}
\centering
\begin{tabular}{lcccc}
\hline
Team  & DSC  & HD95 & Volumetric Similarity & Rank \\ \hline
junma & \textbf{0.41} & 8.96 & \textbf{0.50}                   & 1    \\
joker & 0.40 & \textbf{8.67} & 0.48                   & 2    \\ \hline
\end{tabular}
\end{table}

As shown in Table~\ref{tab:ADAM}, the team `junma' achieved the best DSC and Volumetric Similarity and the team `joker' achieved the best HD95. However, it should be noted that the differences between them are marginal.

\subsection{HECKTOR: 3D Head and Neck Tumor Segmentation in PET/CT}
\begin{figure}[htbp]
\centering
\includegraphics[scale=0.6]{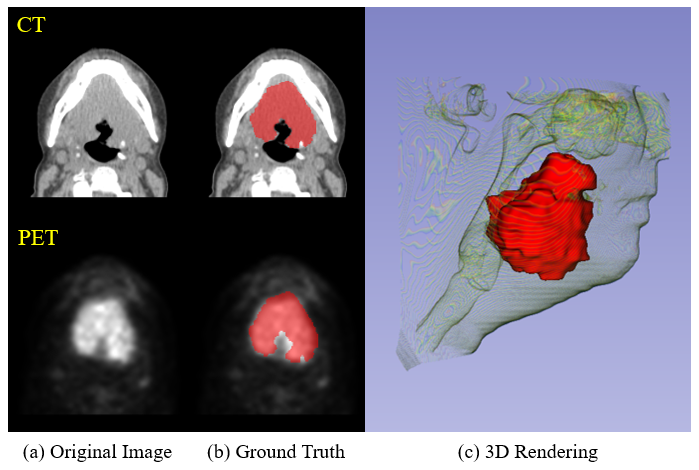}
\caption{Visualized examples in HECKTOR Challenge. Ground truth (b) of the head and neck tumor is shown in 2D projected onto the CT and the PET image (a) and in 3D together with a volume rendering of the raw data (c).
}\label{fig:HECKTOR}
\end{figure}

The task in HECKTOR challenge (\url{http://www.aicrowd.com/challenges/hecktor}) is to segment the heck and neck tumor from PET and CT images.
The organizers provide 201 training cases from four medical centers in Montreal and 53 testing cases from another medical center in Lausane~\cite{HECKTOR-MIDL2020,HECKTOR2021overview}. The tumor ground truth was delineated for radiotherapy treatment planning on PET and CT.
Moreover, the organizers also provided bounding boxes ($114\times114\times114$ $mm^3$) locating the oropharynx region.
The main difficulties are the multi-modality fusion, imbalanced problem, and the unseen testing cases from a new medical center. Figure~\ref{fig:HECKTOR} presents the visualized examples.

Table~\ref{tab:HECKTOR} presents the quantitative segmentation results of the top-2 teams on HECKTOR Challenge Leaderboard\footnote{\url{https://www.aicrowd.com/challenges/miccai-2020-hecktor/leaderboards}} during MICCAI 2020.
Both teams developed their methods based on the two channel 3D U-Net~\cite{cciccek2016UNet3D}. Specifically, the team `andrei.iantsen', replaced the batch normalization with squeeze-and-excitation normalization and introduced the residual blocks in the encoder. The loss function was the unweighted sum of Dice loss and focal loss~\cite{focal2017}.
Four models with leave-one-center-out splits and four additional models with random data splits were trained for 800 epoches using Adam optimizer~\cite{kingma2014adam} on two NVIDIA 1090Ti GPUs with a batch size of 2.
The final model was an ensemble of the eight models.
The team `junma'~\cite{HECKTOR-2020-2nd} firstly trained five 3D nnU-Net models~\cite{isensee2020nnunet} with Dice + TopK loss for five-fold cross-validation. Then, a segmentation quality score was defined by model ensembles, which can be used to select the cases with high uncertainties. Finally, the high uncertainty cases were refined by a hybrid active contour model with iterative convolution-thresholding methods~\cite{wang2017JCP,wang2019ICTM, ma2020ICTM-GAC}.

Both teams concatenated the PET and CT image as input and model ensembles were used to predicting the testing set. In the loss function, Dice loss was also incorporated in both teams.

\begin{table}[!htbp]
\caption{Quantitative results of top-2 teams on HECKTOR Challenge Leaderboard. The bold numbers are the best results.}\label{tab:HECKTOR}
\centering
\begin{tabular}{lcccc}
\hline
Team           & DSC   & Precision & Recall & Rank \\ \hline
andrei.iantsen & \textbf{0.759} & 0.833     & \textbf{0.740}  & 1    \\
junma~\cite{HECKTOR-2020-2nd}  & 0.752 & \textbf{0.838}     & 0.717  & 2    \\ \hline
\end{tabular}
\end{table}

The final rank was based on the DSC scores on the testing set.
The 1st-place team `andrei.iantsen' obtained better DSC and Recall while the 2nd-place team `junma' obtained better Precision. However, the differences between the two teams are marginal, especially for the DSC and Precision.
Moreover, both teams achieved significantly higher Precision than Recall, indicating that most of the segmented voxels were real tumor voxels but many tumor voxels were missed by the model.

\subsection{MyoPS: Multi-sequence CMR based myocardial pathology segmentation challenge}
The task in MyoPS challenge (\url{http://www.sdspeople.fudan.edu.cn/zhuangxiahai/0/myops20/}) is to segment the myocardial pathology (i.e., scar and edema) from multi-sequence cardiac MR images, including the late gadolinium enhancement (LGE) sequence, the T2-weighted sequence, and the balanced- Steady State Free Precession (bSSFP) cine sequence.
The organizers provide 25 cases for training and 20 cases for testing~\cite{MyoPS-MICCAI, MyoPS-TPAMI}.
The main difficulties are the multi-modality fusion, imbalanced problem, low-contrast and heterogeneous appearances of myocardium lesions.

\begin{figure}[htbp]
\centering
\includegraphics[scale=0.6]{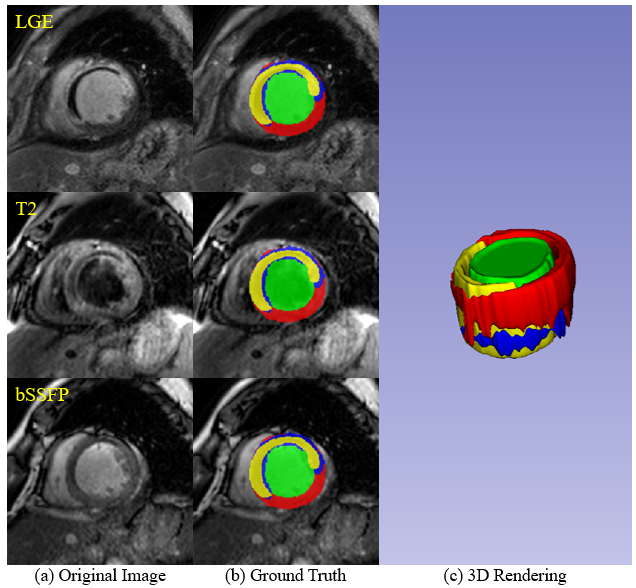}
\caption{Visualized examples in MyoPS Challenge. Ground truth is shown in 2D projected onto the multi-sequence MR images (b) and in 3D rendering (c).
}\label{fig:MyoPS}
\end{figure}

The winner team, `Zhai \& Gu et al.'~\cite{Zhai2020MyoPS}, proposed a coarse-to-fine framework with weighted ensemble. In the coarse segmentation stage, the whole heart was segmented by a U-Net~\cite{ronneberger2015UNet2D} from three sequence MR images.
In the fine segmentation stage, the region of interest (ROI) was cropped according to coarse segmentation results and a nnU-Net~\cite{isensee2020nnunet} was trained to simultaneously segment the left ventricle, right ventricle, healthy myocardium, scar, and edema from the concatenation of three sequence MR images and coarse segmentation results.
Cross-validation results showed that 2D U-Net achieved better performance for the edema while 2.5D U-Net achieved better performance for the scar. To obtained better performance, a weighted method was used for final ensemble. Specifically, the weights for edema and scar prediction channels were 0.8 in 2D and 2.5D U-Net, respectively, while the weights for the other prediction channels were 0.5.

\begin{table}[!htbp]
\caption{Quantitative results of the winner team `Zhai \& Gu et al.'~\cite{Zhai2020MyoPS} in MyoPS challenge.}\label{tab:MyoPS}
\centering
\begin{tabular}{lcc}
\hline
Target       & DSC         & Rank \\ \hline
Scar         & 0.672 $\pm$ 0.244 & 1    \\
Scar + Edema & 0.731 $\pm$ 0.109 & 1    \\ \hline
\end{tabular}
\end{table}

Table~\ref{tab:MyoPS} presents the quantitative segmentation results of the winner team on the testing set\footnote{\url{http://www.sdspeople.fudan.edu.cn/zhuangxiahai/0/myops20/result.html}}. Zhai \& Gu et al. achieved an average DSC of 0.672 $\pm$ 0.244 and 0.731 $\pm$ 0.109 for scar and the combination of scar and edema, respectively. The performance was significantly better than the inter-observer variation of manual scar segmentation (DSC: 0.5243 $\pm$ 0.1578), demonstrating the effectiveness of the proposed method.

\subsection{ABCs: Anatomical Brain Barriers to Cancer Spread: Segmentation from CT and MR Images}
\begin{figure}[htbp]
\centering
\includegraphics[scale=0.75]{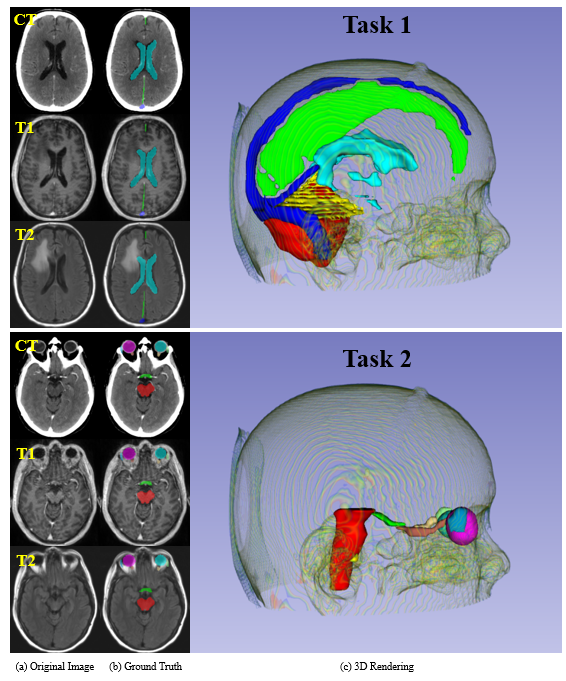}
\caption{Visualized examples in ABCs Challenge. Ground truth (b) is shown in 2D projected onto the multi-modality images (a) and in 3D together with a volume rendering of the raw data (c).
}\label{fig:ABCs}
\end{figure}

ABCs challenge (\url{https://abcs.mgh.harvard.edu/}) included two brain structures segmentation tasks \begin{itemize}
    \item Task 1: segmenting five brain structures, including falx cerebri, tentorium cerebelli, sagittal and transverse brain sinuses, cerebellum and ventricles, which can be used for automated definition of the clinical target volume (CTV) for radiotherapy treatment.
    \item Task 2: segmenting ten structures, including Brainstem, left and right eyes, left and right optic nerves, left and right optic chiasm, lacrimal glands, and cochleas, which can be used in radiotherapy treatment plan optimization.
\end{itemize}
The organizers provide 45 cases for training, 15 cases for validation, and 15 cases for testing. Each case consists of one CT image acquired for treatment planning, and two post-operative brain MRI images (i.e., contrast enhanced T1-weighted, T2-weighted FLAIR). The CT and MR images were obtained from two different CT scanners and seven different MRI scanners, respectively. The multi-modality images were co-registered, and re-sampled to the same resolution and size.
The main difficulties are the multi-modality fusion, imbalanced labels, and multi-vendor cases.
Figure~\ref{fig:ABCs} presents the visualized examples in two tasks.
Participants are required to submit their segmentation results within 48 hours after the time of download the testing set.
DSC and Surface DSC\footnote{\url{https://github.com/deepmind/surface-distance}} at the tolerance of 2 $mm$ are used to evaluate and rank the segmentation results.

\begin{table}[!htbp]
\caption{Quantitative results of top-2 teams on ABCs Challenge Leaderboard. The bold numbers are the best results.}\label{tab:ABCs}
\centering
\begin{tabular}{lccccc}
\hline
\multirow{2}{*}{Team} & \multicolumn{2}{c}{Task 1} & \multicolumn{2}{c}{Task 2} & \multirow{2}{*}{Rank} \\ \cline{2-5}
                      & DSC          & SDSC        & DSC          & SDSC        &                       \\ \hline
Jarvis~\cite{ABCs-2020-1st}                & \textbf{0.888}        & \textbf{0.980}       & \textbf{0.783}        & 0.936       & 1                     \\
HILab                 & 0.883        & 0.978       & 0.781        & \textbf{0.941}       & 2                     \\ \hline
\end{tabular}
\end{table}

Table~\ref{tab:ABCs} presents the average DSC and SDSC of testing set segmentation results of the top-2 teams on the Challenge Leaderboard\footnote{\url{https://abcs.mgh.harvard.edu/index.php/leader-board}}.
Both teams developed their methods based on nnU-Net~\cite{isensee2020nnunet}. Specifically, the team `Jarvis'~\cite{ABCs-2020-1st} used the ResU-Net where residual blocks were introduced in the U-Net encoder.  The training process had three main features:
\begin{itemize}
    \item the training cases `007' and `054' in Task 2 had annotation issues. Thus, the default annotations were replaced with pseudo labels generated by cross-validation.
    \item the flipping along x-axis was dropped from the default data augmentation setting in nnU-Net, because the segmentation targets in Task 2 are sensitive to left and right direction.
    \item in addition to the default Dice + CE loss in nnU-Net, Tversky loss~\cite{salehi2017tversky, SegLossOdyssey} was also used to train the ResU-Net.
\end{itemize}
The final model was an ensemble of default nnU-Net, ResU-Net with Dice-CE loss, and ResU-Net with Tversky loss.
The team `HILab' used a coarse-to-fine framework with nnU-Net~\cite{isensee2020nnunet} for both tasks. Specifically,
\begin{itemize}
    \item in Task 1, an uncovered model was trained to obtain the coarse segmentations with small overfitting. Then, each organ was cropped from the original images and refined by an independent network. The refined organs were combined as the final segmentation results.
    \item in Task 2, a coarse model was firstly trained to segment all organs. Then, each organ was also cropped from the original images and refined by an independent network. The training process was different from Task 1, where a new data augmentation technique, flipping each organ to other side was introduced to enlarge the training set. The final segmentation results were also the combination of the refined organs.
\end{itemize}

Both teams fused the three modalities by concatenating them as the network input. Model ensemble was also used by both teams but the ensemble strategies were different. In particular, the team `Jarvis' used an ensemble of multiple multi-organ segmentation networks while the team `HILab' used an ensemble of one multi-organ and multiple single-organ segmentation networks.

\subsection{BraTS: Brain Tumor Segmentation}
\begin{figure}[htbp]
\centering
\includegraphics[scale=0.5]{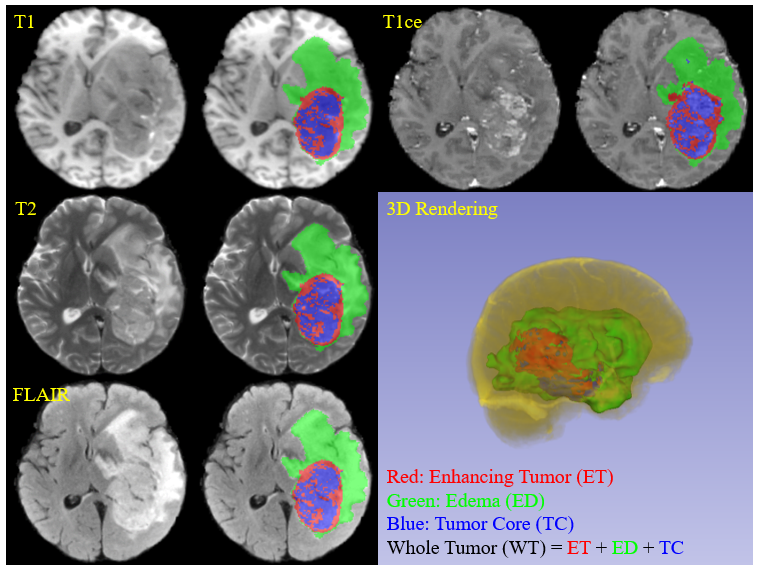}
\caption{Visualized examples in BraTS Challenge. Ground truth is shown in 2D projected onto the multi-sequence MR images and in 3D together with a volume rendering of the raw data.
}\label{fig:BraTS}
\end{figure}

The segmentation task in BraTS challenge (\url{https://www.med.upenn.edu/cbica/brats2020/}) is to segment the enhancing tumor (ET), the tumor core (TC, the necrotic and non-enhancing tumor core), and the whole tumor (WT) from pre-operative multi-modality MR images. As shown in Figure~\ref{fig:BraTS}, the whole tumor comprises the enhancing tumor (red), the edema (green), and the tumor core (blue).
The organizers provide 369 cases for training, 125 cases for testing, and 166 cases for testing. Each case consists of four modalities: the native (T1) MR image, the post-contrast T1-weighted (T1Gd) MR image, the T2-weighted (T2), and the T2 Fluid Attenuated Inversion Recovery (FLAIR) MR image, which were acquired with different clinical protocols and various scanners from 19 institutions~\cite{Brats15-TMI,bratscite3,bratscite4,bratscite5,bakas2018brats}.
The main difficulties are the multi-modality fusion, imbalanced labels, low-contrast and heterogeneous appearances of the brain lesion.
Participants are required to submit their segmentation results within 48 hours after the time of download the testing set.
DSC and HD95 are used to evaluate and rank the segmentation results.

The winner team `MIC-DKFZ', leaded by Fabian et al.~\cite{isensee2020brats-1st}, extended nnU-Net~\cite{isensee2020nnunet} by incorporating BraTS-specific modifications regarding postprocessing, region-based training, a more aggressive data augmentation, BraTS ranking-based model selection as well as several minor modifications, which can improve the default nnU-Net segmentation performance substantially.
Specifically, following BraTS-specific modifications were integrated into nnU-Net’s configuration:
\begin{itemize}
    \item \textbf{Region-based training:} replacing the softmax layer with a sigmoid layer and changing the optimization target to the three tumor sub-regions. The default cross entropy loss term was also replaced with a binary cross-entropy where each of the regions was optimized independently;
    \item \textbf{Postprocesing:} removing enhancing tumor entirely if the predicted volume was less than a given threshold. The threshold was optimized on training set cross-validation twice, once via maximizing the mean Dice score and once via minimizing the BraTS-like ranking score;
    \item \textbf{Increased batch size:} increasing the batch size from 2 to 5;
    \item \textbf{Extensive data augmentation:} using more aggressive augmentations, such as increasing the the probability of applying rotation, scaling, the scale range, elastic deformation, and so on;
    \item \textbf{Batch normalization:} replacing the default  instance normalization with batch normalization;
    \item \textbf{Batch dice:}  computing the dice loss over all samples in the batch;
    \item \textbf{BraTS Ranking-based model selection:} selecting the best model with BraTS-like `rank then aggregate' ranking scheme.
\end{itemize}
The final model was an ensemble of 25 cross-validation models including three groups of top
performing models.

Two tied teams ranked second. Specifically, the team `NPU\_PITT', leading by Jia et al.~\cite{jia2020brats-2nd}, proposed a Hybrid High-resolution and Non-local Feature Network (H$^2$NF-Net) Four modalities were concatenated as a four-channel input and processed at five different scales in the network.
The edema and enhancing tumor were segmented by the single HNF-Net and the tumor core was segmented by he cascaded HNF-Net.
and different brain tumor sub-regions were segmented by single and cascaded HNF-Nets.
The team `Radicals', leading by Wang et al.~\cite{wang2020brats-2nd}, proposed an end-to-end Modality-Pairing learning method with paralleled branches and more layer connections to explore the latent relationship among different modalities. Moreover, a consistence loss was introduced to minimize the prediction variance between branches. The final model was an ensemble of three Modality-Pairing models and three Vanilla nnU-Net~\cite{isensee2020nnunet} models.

\begin{table}[!htbp]
\caption{Quantitative results of the top-3 teams on BraTS 2020 Challenge Leaderboard. The bold numbers are the best results.}\label{tab:BraTS}
\setlength\tabcolsep{3.5pt}
\centering
\begin{tabular}{llcc}
\hline
Team                                                                                                                                                                                               & Target           & DSC                              & HD95                             \\
\hline
\multirow{3}{*}{\begin{tabular}[c]{@{}l@{}}MIC\_DKFZ\\Fabian et al.~\cite{isensee2020brats-1st}\\Rank 1st~\end{tabular}}  & Enhancing Tumor  & 0.820~$\pm$~0.197                & 17.8 $\pm$ 74.9                \\
    & Whole Tumor      & 0.890~$\pm$~0.132 & 8.50 $\pm$ 40.7                \\
                                                                                                                                                                                                   & Tumor Core       & 0.851~$\pm$ 0.240                & \textbf{13.3 $\pm$ 69.5}                \\
\hline
\multirow{3}{*}{\begin{tabular}[c]{@{}l@{}}NPU\_PITT\\Jia et al.~\cite{jia2020brats-2nd}\\Rank 2nd (tie)~\end{tabular}}  & Enhancing Tumor  & \textbf{0.828~$\pm$~0.177}                & \textbf{13.0 $\pm$ 63.7}                \\
    & Whole Tumor      & 0.888~$\pm$~0.119 & \textbf{4.53 $\pm$ 6.21}                \\
                                                                                                                                                                                                   & Tumor Core       & \textbf{0.854~$\pm$ 0.231}                & 16.9 $\pm$ 69.5                \\
\hline
\multirow{3}{*}{\begin{tabular}[c]{@{}l@{}}{Radicals}\\Wang et al.~\cite{wang2020brats-2nd}\\Rank 2nd  (tie) \end{tabular}} & Enhancing Tumor & 0.816 $\pm$ 0.197  & 17.8 $\pm$ 74.9  \\
                                                                                                                                                                                                   & Whole Tumor      & \textbf{0.891 $\pm$ 0.112}  & 6.2 $\pm$ 29.0   \\
                                                                                                                                                                                                   & Tumor Core       & 0.842 $\pm$ 0.244  & 19.5 $\pm$ 74.8  \\
\hline
\end{tabular}
\end{table}

Table~\ref{tab:BraTS} presents the quantitative segmentation results of the top-3 teams on the testing set\footnote{\url{https://www.med.upenn.edu/cbica/brats2020/rankings.html}}. Overall, the performance differences are marginal. The team `MIC\_DKFZ' achieved the best HD95 for the tumor core and the team `Radicals' achieved the best DSC for the whole tumor. The team `NPU\_PITT' achieved the best performance in the remaining four metrics.

\section{Discussion}
\subsection{What are the ``happy-families" elements in the top methods?}
As the Anna Karenina principle goes\footnote{\url{https://en.wikipedia.org/wiki/Anna_Karenina_principle}}:
``All happy families are alike.", there are also some common components in the top methods.

\textbf{nnU-Net~\cite{isensee2020nnunet} backbone} All the top methods used U-Net~\cite{ronneberger2015UNet2D,cciccek2016UNet3D} like architectures in the ten 3D segmentation challenges. Remarkably, nnU-Net was used by the top teams in nine out of ten challenges, because it is open-sourced, powerful, flexible, and out-of-the-box. Participants can easily integrate their new methods into nnU-Net.

\textbf{Dice-related loss functions}
Loss function is one of the most important elements in deep learning-based segmentation methods. nnU-Net used Dice + cross entropy as the default loss function. For extremely imbalanced segmentation tasks, modifying the loss function can obtain better performance.
For example, the winner in HECKTOR challenge used Dice + Focal loss. Both the winner and the runner up used Dice + TopK loss in ADAM challenge. For a more detailed analysis of segmentation loss functions, please refer to \cite{SegLossOdyssey}.

\textbf{Cascaded/coarse-to-fine framework} Cropping the region-of-interest (ROI) can eliminate the unrelated background tissues and reduce the computational burden. Thus, one can firstly trained a model to obtain the coarse segmentation and then crop the ROI. After that, training a new model with the ROI image (concatenated with the coarse segmentation) to refine the segmentation results. This strategy is quite effective for myocardial pathology and small organ segmentation tasks, which was used by both the winners in EMIDEC and MyoPS challenge, and the runner up in ABCs challenge.

\textbf{Model Ensembles} Ensemble is an effective way to fuse the performance of multiple single models. All the top teams used more than one models in their final solutions. The models were usually trained with different data splits, data augmentation techniques, networks, or loss functions, and then combined by averaging the predictions, majority vote, or cascaded frameworks.

\textbf{Concatenated input fusion in multi-modality segmentation tasks}
How to fuse multiple different images is a key question in multi-modality segmentation tasks.
Common deep-learning based image fusion methods include input-level fusion, feature-level fusion, and output-level fusion. In five multi-modality segmentation challenges,
four out of five winner teams used input-level fusion, which directly concatenated multiple images as network inputs. The winner team in ADAM challenge only use one modality but the runner up, achieving similar performance, also used the concatenation strategy to fuse different modalities.

\subsection{Problems and Opportunities}
Based on the summary of the ten segmentation challenges, it can be found that deep learning has achieved unprecedented or even human-level performance on many medical image segmentation tasks, but there still remains several problems.
Following, we introduce some of the problems and also opportunities that can promote the further development of medical image segmentation methods.

\textbf{Standardized method reports}
Many challenge organizers required the participants to submit a short paper to describe their methods. However, these papers are usually structured with their own way and some necessary details might be missed. Currently, the challenge quality has been greatly improved with the Biomedical Image Analysis ChallengeS (BIAS) initiative~\cite{maier2020BIAS-MIA}, where a checklist is used to standardize the review process and raise interpretability and reproducibility of challenge results. Thus, there is also a high demand for the challenge method reports quality control.
The winner team in MICCAI Hackathon Challenge provided an initial attempt
(\url{https://github.com/JunMa11/MICCAI-Reproducibility-Checklist}) at dealing with the method reproducibility with a checklist, but more efforts are required to make this checklist more complete and acceptable by our community.

\textbf{Publicly available baseline models}
nnU-Net has been proved to be a strong baseline.
When starting with a new 3D segmentation challenge, most of the participants will train nnU-Net baseline models, which usually cost 72-120 GPU hours (depending on the computational facilities).
This could be a great waste of energy and time, because participants are repeatedly doing the same thing.
There is a strong demand for publicly available (trained) baseline models when a new segmentation challenge is launched. In this way, participants can pay more attention to developing new methods without spending energy and time on training the baseline models.

\textbf{Fast and memory efficient models}
There is no doubt that accuracy (e.g., DSC, HD) is an important factor for segmentation methods. However, the running time and the GPU memory requirement of the segmentation methods are also critical when deploying the trained models in clinical practice.
Currently, most of the top methods use model ensembles, which could be time-consuming and require very high computing resources.
In order to promote the deep learning-based medical image segmentations to be clinically applicable, it is necessary to pay more attention to the models' running efficiency.

\textbf{Theoretical foundations of segmentation models}
Current theoretical studies of deep learning usually have strong assumptions~\cite{CNN-Theory-PNAS,E-CNN-Theory,CNN-Theory-Tao}, such as smoothness, infinite width, and so on. However, when it comes to real practice, many open problems remain unsolved. For example, what is the theoretical principle of designing segmentation network architectures? is there a generalization gap? how should we to estimate it?
what does the loss function landscape look like?
does the training process converge to a good solution? How fast?
how much data do we need when start with a new segmentation task?

\textbf{Diverse datasets and generalizable segmentation models}
Collecting diverse datasets is critical for developing generalizable segmentation models, because clinical practice requires the trained models can be applied to many (unseen) medical centers.
According to the challenge results (e.g., M\&Ms, BraTS, HECKTOR), we found that the segmentation performances have a significant drop when testing sets include unseen cases from new medical centers.
Thus, it is important to have diverse datasets to evaluate the models' generalization ability when organizing segmentation challenges.
Currently, building generalizable models that can be applied consistently across medical centres, diseases, and scanner vendors is still an unsolved and challenging problem.

\textbf{Extremely imbalanced target segmentation}
Imbalanced segmentation has been a long-term problem in medical image segmentation, especially when the size of target foreground region is several orders of magnitude less than the background size.
Recently studies have made some progress \cite{milletari2016vnet,ma-MIDL2020-SegWithDist,BDLoss-2020-MIA}, however, the extremely imbalanced segmentation still remains very difficult. For example, in ADAM challenge, the median target size is 238, which occupies $6.5\times10^{-6}$ of the median image size.
The winner method achieved a DSC score of 0.41, which has large rooms for further improvements.

\subsection{Limitations}
There are also some important but not mentioned topics in this paper.
For example, a summary of 2D medical image segmentation challenges~\cite{MoNuSeg,ross2020ROBUST-MIS,AGE-Challenge} has not been included in this paper, because we only found three 2D international image segmentation challenges in 2020, including thyroid nodule segmentation in ultrasound images (\url{https://tn-scui2020.grand-challenge.org/Home/}), optic disc and cup segmentation in fundus images (\url{https://refuge.grand-challenge.org/Home2020/}), and cataract segmentation in surgical videos (\url{https://cataracts-semantic-segmentation2020.grand-challenge.org/Home/}).
Thus, the findings would be biased with limited challenge samples and we will provide a similar summary for 2D medical image segmentation methods when there are many ($\geq10$) international challenges. Moreover, this paper only covers cutting-edge fully supervised segmentation methods, while semi-supervised learning~\cite{Not-so-supervised-MIA-19,van2020semi-survey, Luo2020smalldata-survey}, weakly-supervised learning~\cite{tajbakhsh2020embracing-MIA,NoisyLabel-Review}, and continual learning~\cite{ContinualLearning-NN,soltoggio2018borntolearn,hoi2018onlineLearning-Survey}-based segmentation methods are not mentioned. This is because, currently, there are few benchmarks or challenges~\cite{ma2020COVID-Data,Ma-2020-abdomenCT-1K} for these topics in the medical image segmentation field.

\section{Conclusion}
Challenges provide an open and fair platform for various research groups to test and validate their segmentation methods on common datasets acquired from the clinical environment.
In this paper, we have summarized ten 3D medical image segmentation challenges and the corresponding top methods.
In addition, we also identify the widely involved "happy-families" elements in the top methods and give potential future research directions in medical image segmentation.
Moreover, we also maintain a public GitHub repository (\url{https://github.com/JunMa11/SOTA-MedSeg}) to collect the cutting-edge segmentation methods based on various international segmentation challenges.
We expect that this review of the cutting-edge 3D image segmentation methods will be beneficial to both early-stage and senior researchers in related fields.


%



\ifCLASSOPTIONcompsoc
  \section*{Acknowledgments}
\else
  \section*{Acknowledgment}
\fi

The authors would like to thank all the organizers for creating the public datasets and holding the great challenges.
The authors also highly appreciate Ruochen Gao (the winner in ASOCA), Ran Gu (the winner in MyoPS), Wenhui Lei (the winner in MyoPS and the runner up to winner in ABCs), Munan Ning (the winner in ABCs), and Yixin Wang (the runner up to winner in BraTS), Yao Zhang (the runner up to winner in M\&Ms), Yichi Zhang (the winner in EMIDEC) for valuable discussions.

\ifCLASSOPTIONcaptionsoff
  \newpage
\fi

\bibliographystyle{IEEEtran}
\bibliography{JunRef}




\end{document}